# Strong interfacial exchange field in a heavy metal/ferromagnetic insulator system determined by spin Hall magnetoresistance

Juan M. Gomez-Perez,[1] Xian-Peng Zhang,[3,4] Francesco Calavalle,[1] Maxim Ilyn,[4] Carmen González-Orellana,[4] Marco Gobbi,[1,2,4] Celia Rogero,[3,4] Andrey Chuvilin,[1,2] Vitaly N. Golovach,[2,3,4,5] Luis E. Hueso,[1,2] F. Sebastian Bergeret,[3,4] Fèlix Casanova[1,2,*]

[1]CIC nanoGUNE BRTA, 20018 Donostia-San Sebastian, Basque Country, Spain
[2]IKERBASQUE, Basque Foundation for Science, 48013 Bilbao, Basque Country, Spain
[3]Donostia International Physics Center, 20018 Donostia-San Sebastian, Basque Country, Spain
[4]Centro de Física de Materiales CFM-MPC (CSIC-UPV/EHU), 20018 Donostia-San Sebastian, Basque Country, Spain
[5]Departamento de Física de Materiales UPV/EHU, 20018 Donostia-San Sebastian, Basque Country, Spain

**Abstract:** Spin-dependent transport at heavy metal/magnetic insulator interfaces is at the origin of many phenomena at the forefront of spintronics research. A proper quantification of the different interfacial spin conductances is crucial for many applications. Here, we report the first measurement of the spin Hall magnetoresistance (SMR) of Pt on a purely ferromagnetic insulator (EuS). We perform SMR measurements in a wide range of temperatures and fit the results by using a microscopic model. From this fitting procedure we obtain the temperature dependence of the spin conductances ($G_s$, $G_r$ and $G_i$), disentangling the contribution of field-like torque ($G_i$), damping-like torque ($G_r$), and spin-flip scattering ($G_s$). An interfacial exchange field of the order of 1 meV acting upon the conduction electrons of Pt can be estimated from $G_i$, which is at least three times larger than $G_r$ below the Curie temperature. Our work provides an easy method to quantify this interfacial spin-splitting field, which play a key role in emerging fields such as superconducting spintronics and caloritronics, and topological quantum computation.

Spin transport in systems consisting of magnetic insulators (MIs) and non-magnetic metals is of extreme importance in the field of spintronics. The spin currents through the interface of such heterostructure are at the origin of many phenomena, from spin pumping[1] to spin Seebeck effect[2], or spin Hall magnetoresistance[3-32]. The spin transport at the interface can be described in terms of three parameters: the spin-sink conductance $G_s$, which originates when the electron spins of the non-magnetic metal are collinear with the MI magnetization[33-35], and the real and imaginary part of the spin-mixing conductance, $G_{\uparrow\downarrow}= G_r + iG_i$ (refs 36 and 37), which originate from torques that the electron spins of the non-magnetic metal exert to the magnetization of the MI when they are noncollinear. $G_r$ is determined by the Slonczewski (or damping-like) torque, an important quantity for current-induced magnetization switching in spin-transfer torque magnetic random-access memory (STT-MRAM) devices, currently ready for mass production[38], as well as in spin-orbit torque devices[16,39]. On the other hand, $G_i$ quantifies the exchange field between the electrons of the non-magnetic metal and the magnetic moments of the MI, exerting a field-like torque when spin accumulation is induced. This interfacial exchange field is very relevant in different areas. For instance, when the non-magnetic metal is a superconductor, it leads to a spin-splitting field, even in the absence of an external magnetic field[40-43]. Such spin-splitting in superconductors are subjected to intense research[44,45] because of their possible applications in cryogenic memories[46], thermoelectric detectors[47], superconducting spintronics and caloritronics[48,49], and in the field of topological superconductivity induced in superconducting hybrid structures[50,51]. Such exchange field has also been used to induce ferromagnetism in graphene[52,53].

Spin Hall magnetoresistance (SMR) is a simple, yet powerful technique that can be used to quantify the interfacial spin conductances. When a heavy metal (HM), with a sizable spin Hall effect, is placed in contact with a MI, the SMR appears as a modulation of the HM resistivity, governed by $G_r$–$G_s$ (ref 54), which follows the relative orientation between the magnetization ($\boldsymbol{M}$) in the MI and the spin-Hall induced spin accumulation ($\boldsymbol{\mu}_s$) in the HM. SMR has been extensively studied in different MIs, for instance, ferrimagnetic insulators such as $Y_3Fe_5O_{12}$ (refs 3-15), $Tm_3Fe_5O_{12}$ (refs 16 and 17), $Gd_3Fe_5O_{12}$ (compensated ferrimagnet)[18] or $Cu_2OSeO_3$ (spiral ferrimagnet)[19], antiferromagnetic insulators such as NiO, $Cr_2O_3$ and CoO (refs 20-26), low dimensional ferromagnets[27] or even paramagnetic insulators[28-31]. SMR also shows up as an anomalous Hall-like contribution in the HM, in this case governed by $G_i$ (refs 15, 32 and 54). In the studied cases with ferrimagnetic garnets, $G_r$ is at least one order of magnitude larger than $G_i$ (refs 11-13, 16, 17, 37). In 2006, A. Brataas et al.[36] suggested that, at the interface with a ferromagnetic insulator (FMI), such as an europium chalcogenide, $G_i$ might dominate over $G_r$, a prediction also pointed out more recently by some of the authors from a microscopic model of SM[54]. However, up to date, there are no reports characterizing SMR in a purely FMI, mainly because of the small number of FMIs with large enough Curie temperature ($T_c$) available.

In this letter, we report the SMR in a HM such as Pt on top of EuS, a FMI with a $T_c$ around 19 K. The temperature dependence of the SMR amplitudes can be fit using a model of ferromagnetism with the microscopic theory for SMR.[54] From the fittings, we can quantify the exchange interaction between the conduction electrons of Pt and the localized moments of Eu (~3–4 meV), as well as the values of $G_s$, $G_r$ and $G_i$ as a function of temperature. We demonstrate that, in a FMI where there is no compensation of magnetic moments at the interface, $G_i$ is larger than $G_r$. The precise quantification of the interfacial exchange field from $G_i$ is relevant in many fields where this quantity plays a crucial role.

A detailed description of the fabrication process and characterization of the EuS/Pt samples is given in methods. The evaporated EuS shows an insulating behavior (see Note S1). Figure 1a shows the temperature dependence of the EuS magnetization measured by vibrating sample magnetometry (VSM) at constant magnetic field ($\mu_0 H = 0.02$ T). EuS exhibits a ferromagnetic behavior with a broad transition at $T_c$ ~19 K, in agreement with previous reports[55,56], but $M$ does not saturate down to 2.5 K, a trend observed in similarly evaporated EuS films[57]. The in-plane $M(H)$ curve at 5 K (inset of Figure 1a) shows a clear hysteresis loop, with a coercive field around ~3 mT, and saturates at ~20 mT, confirming the soft magnetic behavior of the EuS film. Our magnetization saturation is below its bulk value[58], as expected in thin films[59] (see Note S2 for further discussion). Importantly for the SMR measurements, the TEM analysis (Figure 1b) shows the good quality of the interface between the Pt and EuS, where EuS has the right composition and crystallographic structure (fcc), which is grown textured in the (200) orientation (see Note S3 for details).

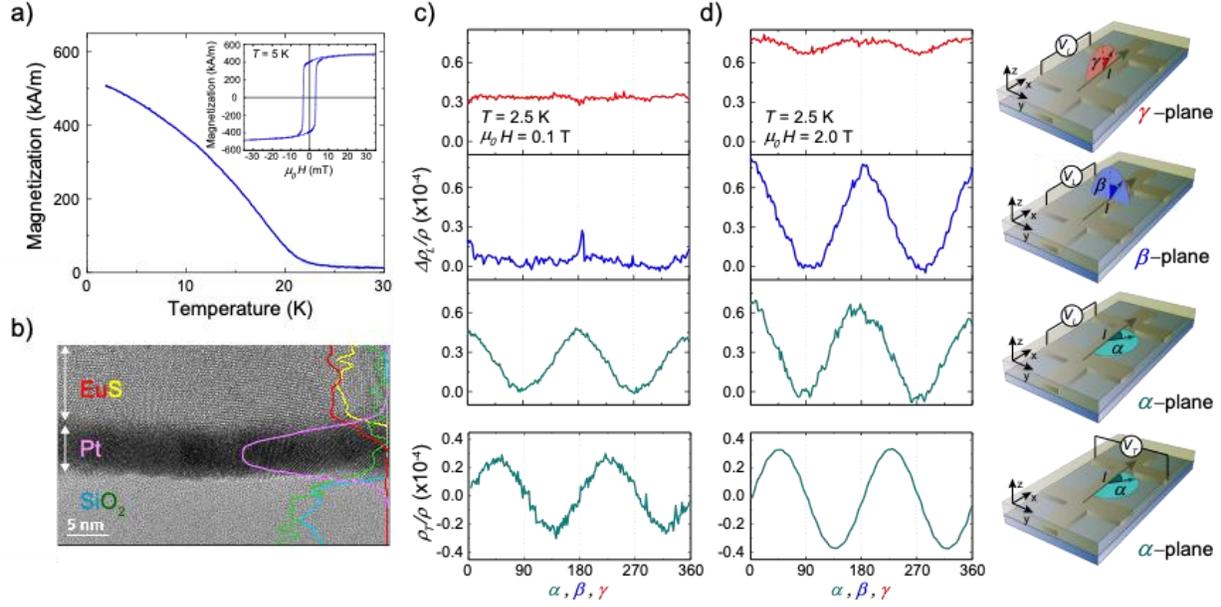

**Figure 1.** (a) Temperature dependence of the EuS magnetization measured at $\mu_0 H = 0.02$ T. Inset: Magnetic hysteresis loop performed at 5 K. (b) TEM image of the $SiO_2$/Pt/EuS heterostructure. The solid lines on the right side represent the composition profile along the sample thickness obtained by EDX. The colors correspond to Si (blue), O (green), Pt (purple), Eu (red), and S (yellow). (c-d) Normalized longitudinal ($\Delta\rho_L/\rho$) and transverse ($\rho_T/\rho$) ADMR measurements at 2.5 K along the three relevant $H$-rotation planes ($\alpha$, $\beta$, $\gamma$) (see sketches on the right side) for different applied magnetic fields: (c) $\mu_0 H = 0.1$ T and (d) $\mu_0 H = 2$ T, respectively.

According to the original SMR theory[32], the longitudinal and transverse resistivity in a HM layer in contact with a MI depends on the direction of the MI magnetization as follows:

$$\rho_L = \rho + \Delta\rho_0 + \Delta\rho_1(1-m_y^2),$$
$$\rho_T = \Delta\rho_1 m_x m_y + \Delta\rho_2 m_z, \qquad (1)$$

where $\mathbf{m} = (m_x, m_y, m_z)$ is a unit vector along $\mathbf{M}$, $\rho$ is the Pt resistivity, $\Delta\rho_0$ accounts for a resistivity correction due to the spin Hall effect, and $\Delta\rho_1$ and $\Delta\rho_2$ are the SMR amplitudes.

The longitudinal ($R_L = V_L/I$) and transverse ($R_T = V_T/I$) resistance are measured using the standard configurations shown in the sketches on the right side of Figure 1. Figures 1c and 1d show the longitudinal ($\Delta\rho_L/\rho$) and transverse ($\rho_T/\rho$) angular dependent magnetoresistance (ADMR) measurements at 2.5 K with the magnetic field ($\mu_0 H = 0.1$ T and 2 T, respectively) rotating along three main planes ($\alpha$, $\beta$, and $\gamma$) defined in the sketches on the right side of the panels. $\Delta\rho_L/\rho$ and $\rho_T/\rho$ are obtained as $\Delta\rho_L/\rho = [R_L(\alpha,\beta) - R_L(90°)]/R_L(90°)$, where $R_L(90°) = 170\ \Omega$, and $\rho_T/\rho = d_N(L/w)[R_T(\alpha) - R_T(90°)]/\rho$, where $\rho$ is $\sim 47\ \mu\Omega\cdot$cm and $d_N$ is the thickness of the HM. At $\mu_0 H = 0.1$ T (Figure 1c), only the longitudinal (transverse) ADMR in $\alpha$–plane follows the $cos^2$ ($cos\cdot sin$) dependence predicted by eqs 1 (refs 3, 6, 9, 11, 12 and 27), because only the in-plane magnetization follows the external field, as expected from the shape anisotropy of a soft ferromagnet such as EuS. In order to saturate the EuS film out-of-plane, we applied $\mu_0 H = 2$ T (Figure 1d). In this case, a clear $cos^2$ dependence of the longitudinal ADMR in $\alpha$– and $\beta$–plane and a *cos·sin* dependence of the transverse ADMR in $\alpha$–plane is observed, all with a similar amplitude, which corresponds to $\Delta\rho_1/\rho$. This symmetry follows well eqs 1, and is linked to the interaction between $\boldsymbol{\mu}_s$ induced by the spin Hall effect[60,61] in the Pt when a charge current is applied and the magnetic moments of the EuS at the interface. When $\boldsymbol{\mu}_s$ and the magnetic moments are parallel, the spins are reflected at the interface and converted back into a charge current by the inverse spin Hall effect, decreasing the overall Pt resistance. However, when $\boldsymbol{\mu}_s$ and the magnetic moments are perpendicular, $\boldsymbol{\mu}_s$ exerts a torque to $\boldsymbol{M}$ and part of the spin angular momentum is absorbed by the MI, resulting in an increase of the Pt resistance. In the case of $\gamma$–plane, no modulation is expected from eqs 1. Nonetheless, there is a small modulation not related to SMR, which is identified as weak antilocalization (WAL) that appears in Pt at low temperatures and large out-of-plane field[62,63]. An anisotropic magnetoresistance (AMR) origin of this modulation due to magnetic proximity effect (MPE) in Pt is ruled out (see Note S4 for details).

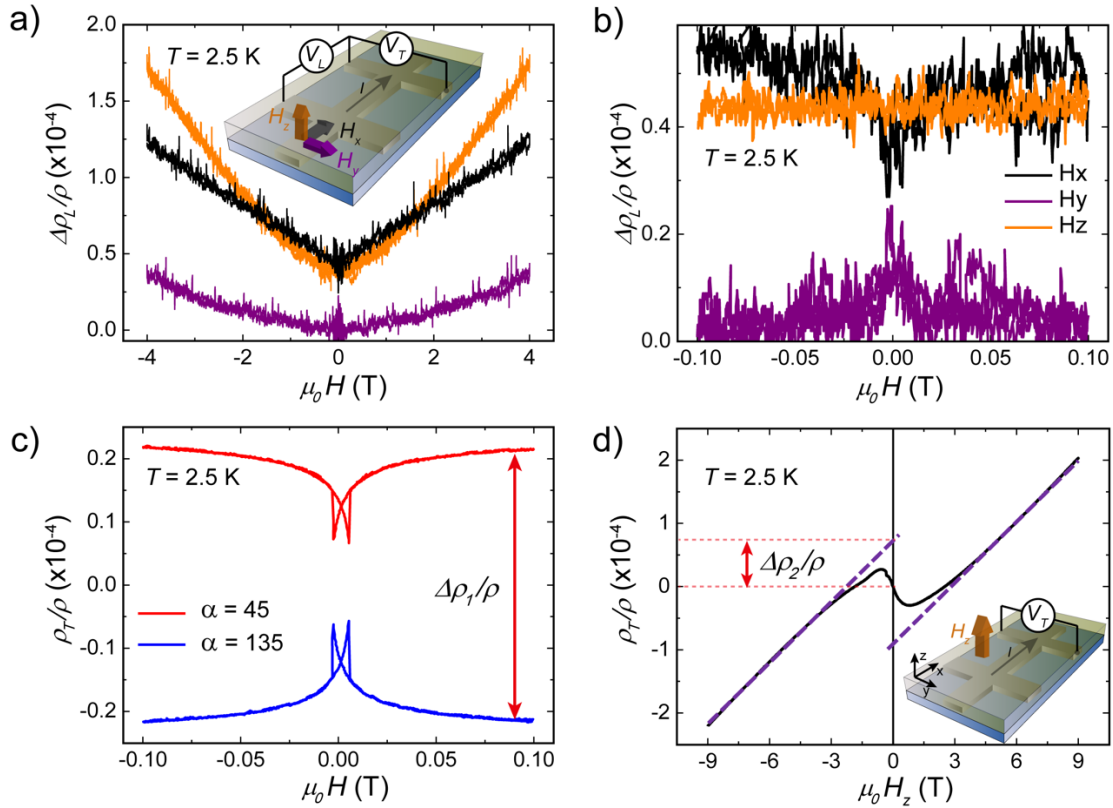

**Figure 2.** (a) Longitudinal FDMR measurements performed along the three main axes at 2.5 K in a range of magnetic fields between 4 T and –4 T (see sketch for the definition of the axes, color code of the magnetic field direction, and measurement configuration). (b) Zoom of panel a at low magnetic fields between 0.1 and -0.1 T, where the magnetization reversal occurs. (c) Transverse FDMR measurements performed with the applied magnetic field in plane at $\alpha = 45°$ and $\alpha = 135°$. The red arrow shows the amplitude corresponding to $\Delta\rho_1/\rho$. (d) Transverse resistivity measurement in Hall configuration (see sketch for the measurement configuration). Dash purple lines correspond to the linear fit performed at large magnetic fields and extrapolated to zero. The red arrow shows the amplitude corresponding to $\Delta\rho_2/\rho$.

The observation of SMR in our system is confirmed by the longitudinal field dependent magnetoresistance (FDMR) measurements shown in Figure 2b in the three main axes (see sketch in Figure 2a). The curve with $H_z$ applied should be the same as the one with $H_x$ due to the spin symmetry by SMR; in both cases, $\boldsymbol{\mu}_s$ is perpendicular to the EuS magnetization, leading to a high resistance state. The curve with $H_y$ applied shows lower resistance than the ones with $H_x$ and $H_z$ as expected from SMR. At higher magnetic fields, the FDMR with $H_z$ is larger than with $H_x$, which is in turn larger than with $H_y$ (see Figure 2a). This behavior arises from a combination of WAL and Hanle magnetoresistance (HMR)[63] in Pt, as detailed in Note S4. At low magnetic fields, a clear gap corresponding to the $\Delta\rho_1/\rho$ amplitude appears between the curve with $H_x$ and the curve with $H_y$, with peaks around zero field corresponding to the reversal magnetization of EuS (~4 mT in plane) (see Figure 2b). The same information can be obtained, with larger signal to noise ratio, from the transverse resistivity. Figure 2c plots the transverse FDMR, with the magnetic field applied in plane, at angles $\alpha = 45º$ and $\alpha = 135º$ (corresponding to the maximum and minimum resistance values), confirming both the $\Delta\rho_1/\rho$ amplitude and the magnetization reversal around ~4 mT. The coercive field observed in the FDMR measurements decreases with temperature (~3 mT at 5 K, ~1.4 mT at 10 K, not shown). The value at 5 K is in perfect agreement with the $M(H)$ value obtained from the inset of Figure 1a, confirming that SMR is sensitive to the magnetization orientation of the EuS film. For details on the low-field behavior of the curve with $H_z$, see Note S5.

Next, we performed measurements with a Hall configuration (i.e., transverse resistivity with $H_z$, see sketch in Figure 2d). At large $H_z$, we observe a linear dependence of $\rho_T$ with $H_z$ that corresponds to the ordinary Hall effect in the Pt. At low $H_z$, $\rho_T$ shows an anomalous Hall-like feature which follows the out-of-plane magnetization reversal of EuS. This observed feature corresponds to the $\Delta\rho_2$ term of the SMR theory (eqs 1). $\Delta\rho_2$ appears when the magnetization is out-of-plane since $\boldsymbol{\mu}_s$ precesses around $m_z$, an effect quantified by $G_i$. Since the spin polarization rotates from the $y-$ to the $x-$direction, the inverse spin Hall effect acting in the Pt converts the spins back to charge current along the $y-$direction, leading to the measured transverse voltage[32]. An anomalous Hall effect origin caused by MPE in Pt is ruled out by the previous ADMR and FDMR measurements, which do not show trace of AMR in Pt. The SMR origin of this anomalous Hall-like feature has been recently confirmed by using Au instead of Pt[11]. $\Delta\rho_2/\rho$ can be extracted from Figure 2d with the intercept of the linear fittings at large (positive and negative) magnetic fields. According to the original SMR theory[32], $\Delta\rho_1/\rho$ essentially depends on the real part of the spin-mixing conductance, whereas $\Delta\rho_2/\rho$ mostly depends on the imaginary part. The anomalous Hall-like amplitude, which has only been reported few times in ferrimagnetic garnets[11-13,16,17,37], is small because $G_i$ is at least one order of magnitude smaller than $G_r$ in these ferrimagnetic insulators due to the partial compensation of the magnetic moments at the surface. Nonetheless, in our EuS/Pt system, we can see a clear $\Delta\rho_2/\rho$ amplitude. Figure 3b plots the temperature dependence of both $\Delta\rho_1/\rho$ (extracted at 0.1 T, above the magnetization saturation but low enough to avoid any HMR contribution) and $\Delta\rho_2/\rho$, showing that $\Delta\rho_2/\rho$ is larger, in absolute value, than $\Delta\rho_1/\rho$. Both amplitudes disappear close to the $T_c$ of the EuS film.

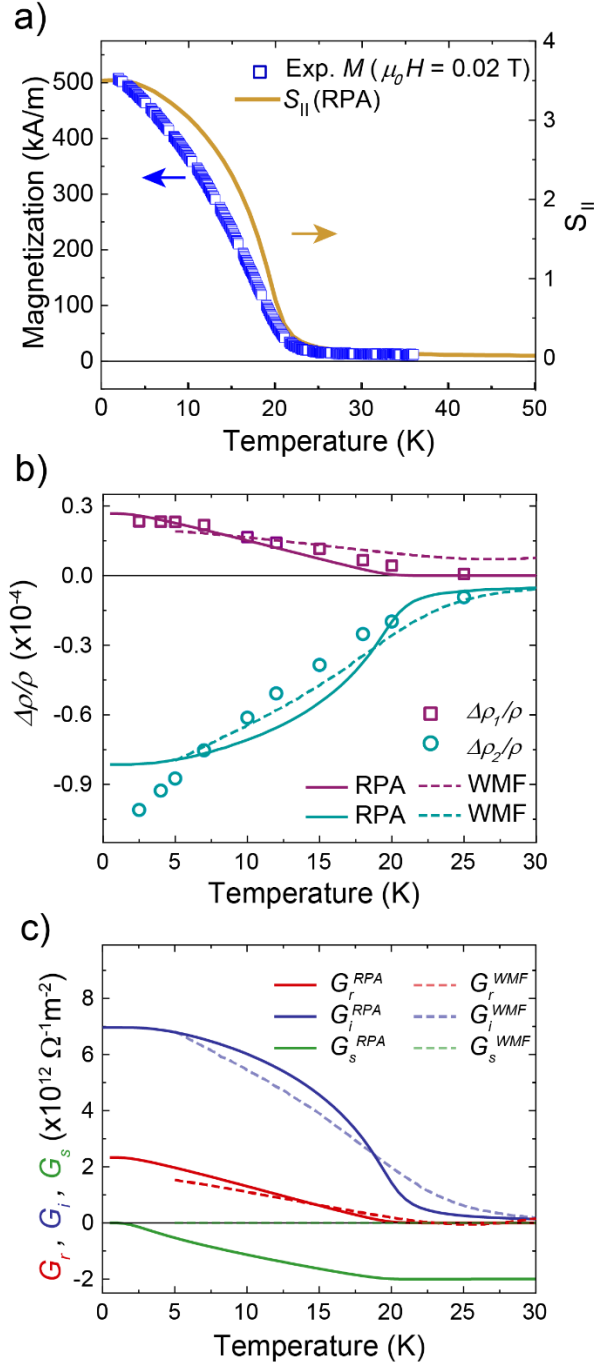

**Figure 3.** (a) Temperature dependence of the experimental EuS magnetization (blue open squares) and of the spin-operator $S_\parallel$ (dark yellow line) extracted from the fitting of the experimental curve to the RPA model for ferromagnetism. (b) Temperature dependence of the normalized SMR amplitudes $\Delta\rho_1/\rho$ (extracted from the transverse FDMR at $\mu_0 H = 0.1$ T as the example shown in Figure 2c) and $\Delta\rho_2/\rho$ (extracted from the Hall configuration measurement as the example shown in Figure 2d). The open dots represent the experimental data and the solid (dashed) lines are the amplitudes obtained with the RPA (WMF) model and the microscopic theory[54]. (c) Temperature dependence of the real part ($G_r$), imaginary part ($G_i$) of the spin-mixing conductance and the spin-sink conductance ($G_s$). The solid (dashed) lines are calculated values from the best fits obtained with the RPA (WMF) model.

In the original SMR theory[12], the spin-mixing conductance terms are assumed to be temperature and field independent. The spin current at the HM/MI interface is given by[64,65]

$$-e\mathbf{J}_{s,z} = G_s \boldsymbol{\mu}_s + G_r \mathbf{n} \times [\mathbf{n} \times \boldsymbol{\mu}_s] + G_i \mathbf{n} \times \boldsymbol{\mu}_s, \quad (2)$$

where $e$ is the elementary charge, $\mathbf{J}_{s,z}$ is the spin current flowing in $z$ direction, $\boldsymbol{\mu}_s$ the vector of the spin accumulation and $\mathbf{n}$ is the unit vector in the direction of the applied magnetic field $H$. $G_r$ and $G_i$ are the spin-mixing conductance terms. The origin of $G_i$ is the interfacial exchange field which is an equilibrium property. When the HM is driven out of equilibrium by passing a charge current, the exchange field interacts with the induced spin accumulation, which is described by the last term in Eq. (2). The spin-sink conductance $G_s$ is related to the spin-flip scattering at the interface and plays an important role for the excitation of magnons in MIs[66–69]. It is important to emphasize that, in order to write eq 2 in this customary form, $G_s$ is a negative quantity (see ref 54 for more details). Since we are interested in the temperature dependence of the interfacial spin conductances, it is necessary to use a microscopic derivation of these parameters. As shown in ref 54, the different spin conductances can be defined in terms of the spin-spin correlation functions of the local magnetic moments at the surface of the MI. This allows to study their temperature and magnetic field dependencies[54]. Hence, $G_r$, $G_i$ and $G_s$ are defined as

$$G_r = e^2 v_F \left(\frac{1}{\tau_\perp} - \frac{1}{\tau_\parallel}\right),$$
$$G_i = -\frac{e^2}{\hbar} n_{imp}^{2D} v_F J_{int} \langle \hat{S}_\parallel \rangle,$$
$$G_s = -e^2 v_F \frac{1}{\tau_\parallel}, \quad (3)$$

with

$$\frac{1}{\tau_\parallel} = \frac{2\pi}{T} n_{imp}^{2D} v_F J_{int}^2 \omega_L^m n_B(\omega_L^m)[1 + n_B(\omega_L^m)] |\langle \hat{S}_\parallel \rangle|,$$
$$\frac{1}{\tau_\perp} = \frac{1}{2\tau_\parallel} + \frac{\pi}{\hbar} n_{imp}^{2D} v_F J_{int}^2 \langle \hat{S}_\parallel^2 \rangle, \quad (4)$$

where $T$ is the temperature, $n_B = 1/(e^{\hbar\omega/k_B T} - 1)$ is the Bose-Einstein distribution function with $k$ the Boltzmann constant, $\omega_L^m = \omega_B - \langle \hat{S}_\parallel \rangle \sum_j J_{ij}/\hbar$ with $J_{ij}$ being the coupling constant of the Heisenberg ferromagnet, $\omega_B = g\mu_B B/\hbar$ with $g$ the gyromagnetic factor and $\mu_B$ the Bohr magneton, $v_F$ is the density of the electronic states per spin species in the HM at the Fermi level, $J_{int}$ is the exchange interaction between the electron spins in the Pt and the magnetic moments of the FMI, $n_{imp}^{2D}$ is the surface density of localized magnetic moments at the interface, and $\hbar$ the reduced Planck constant. $\hat{S}_\parallel$ is the longitudinal spin operator of a representative local moment, with $\langle \hat{S}_\parallel \rangle$ being the spin expectation value, which is the projection of the localized moment parallel to $H$. In addition, $\langle \hat{S}_\parallel^2 \rangle$ is the spin-spin correlation function obtained from $\hat{S}_\parallel$. Due to the spontaneous magnetization in a FMI, the spin relaxation times become anisotropic ($\tau_\parallel \neq \tau_\perp$) leading to the appearance of SMR. In this scenario, the SMR amplitudes $\Delta\rho_1/\rho$ and $\Delta\rho_2/\rho$ are given by

$$\Delta\rho_1/\rho = \theta_{SH}^2 \{\mathcal{F}(G_s, \lambda) - \Re[\mathcal{F}(G_s - G_{\uparrow\downarrow}, \Lambda)]\},$$
$$\Delta\rho_2/\rho = \theta_{SH}^2 \Im[\mathcal{F}(G_s - G_{\uparrow\downarrow}, \Lambda)], \quad (5)$$

where $\frac{1}{\Lambda} = \sqrt{\frac{1}{\lambda^2} + i\frac{1}{\lambda_m^2}}$ with $\lambda_m = \sqrt{\frac{D\hbar}{g\mu_B|B|}}$, $D$ the diffusion coefficient of the HM, $\lambda$ the spin diffusion length of the HM, and $\theta_{SH}$ the spin Hall angle of the HM. The auxiliary function $\mathcal{F}(G,\lambda)$ is defined as

$$\mathcal{F}(G,\lambda) = \frac{2\lambda}{d_N} \tanh\left(\frac{d_N}{2\lambda}\right) \frac{1-\rho G\lambda \coth\left(\frac{d_N}{2\lambda}\right)}{1-2\rho G\lambda \coth\left(\frac{d_N}{\lambda}\right)}. \tag{6}$$

In order to fit simultaneously the temperature dependence of the two SMR amplitudes, we need to model the magnetic ordering in EuS. For this, we used two possible approaches: the random phase approximation (RPA) and the Weiss mean field theory (WMF). First of all, by using the RPA model, we fit the experimental magnetization (Figure 3a) to extract the exchange integral for first neighbors, $J_1$ (12 first neighbors), and the second neighbors, $J_2$ (6 second neighbors) for 4$f$ electrons in EuS. The values extracted from the model fit are $0.27k_B$ for $J_1$ and $-0.12k_B$ for $J_2$, which match well the experimental ones obtained by neutron scattering, $J_1 = 0.221k_B$ and $J_2 = -0.100k_B$ (refs 70 and 71). In the RPA model, we calculate $\langle \hat{S}_\parallel \rangle$ and $\langle \hat{S}_\parallel^2 \rangle$ from the obtained $J_1$ and $J_2$ values, whereas in the WMF model, we use the experimental magnetization as $\langle \hat{S}_\parallel \rangle$, and from $\langle \hat{S}_\parallel \rangle$, we calculate $\langle \hat{S}_\parallel^2 \rangle$ (ref 54). As for the Pt parameters, we take the values of $\lambda$ (~1.3 nm) and $\theta_{SH}$ (~0.19) corresponding to the Pt resistivity at each temperature[72]. Figure 3b shows the fits of the SMR amplitudes $\Delta\rho_1/\rho$ (purple lines) and $\Delta\rho_2/\rho$ (blue lines) to the RPA (solid lines) and WMF (dashed lines) models, with $J_{int}$ and $n_{imp}^{2D}$ being the only free parameters. The fits reproduce reasonably well the two experimental curves for both the RPA and WMF models. The parameter values obtained in the fittings are: i) $J_{int}/a^3 = 4.3$ meV (RPA) and 3.0 meV (WMF), which is the exchange interaction between the conduction electrons in Pt and the localized magnetic moments in EuS and is ferromagnetic ($J_{int}/a^3 > 0$); (ii) $n_{imp}^{2D} = 0.10/a^2$ (RPA) and $0.14/a^2$ (WMF), with $a = 5.94$ Å being the EuS lattice parameter. The best fitting parameters are similar for both methods, which strengthen the reliability of the obtained values. The obtained $J_{int}$ here is of the same order as the ~10 meV experimentally obtained for YIG/Pt (ref [73]). $n_{imp}^{2D}$ values are 10–14% of the ideal value, but they depend on the slicing of the lattice surface and the quality of the HM/FMI surface. The difference between the RPA model and the experimental data arises from the fact that EuS does not behave as an ideal ferromagnet and the $M(T)$ curve cannot be fully captured by a simple model (see Figure 3a). In this regard, the WMF model gives a better fit because it uses the experimental $M(T)$ curve to extract one of the required parameters ($\langle \hat{S}_\parallel \rangle$).

From the same fitting parameters obtained from the fits in Figure 3b, we use eqs 3 to calculate $G_r$, $G_i$, and $G_s$ as a function of temperature, which are plotted in Figure 3c. At low temperatures, when the EuS magnetization saturates, $G_r$ and $G_i$ are largest since $\langle \hat{S}_\parallel \rangle$, and thus the torque, is maximum. However, because the magnetization is frozen, the conservation of the angular momentum leads to a reduction of the spin-flip scattering and hence to a suppression of $G_s$. In contrast, at higher temperatures and close to $T_c$, with the absence of net magnetization due to the randomized spins ($\langle \hat{S}_\parallel \rangle \to 0$), $G_r$ and $G_i$ vanish to zero because of the isotropic relaxation time (see eqs 5), whereas $|G_s|$ becomes maximum. A key point of the obtained results is that we experimentally demonstrate for the first time that, in a FMI such as EuS, $G_i$ is larger than $G_r$, at least 3 times larger at the lowest measured temperature (2.5 K), as predicted in ref 36 for europium chalcogenides. According to our results, we can confirm that the field-like torque plays an important role in Pt/EuS because all magnetic moments of the interface contribute to the interfacial exchange field, as opposed to the ferrimagnetic case[11-13,16,17,37], where there is a compensation of the magnetic moments. Quantitatively, in the strongly magnetized regime, we have $G_r = G_Q S^2$ and $G_i = G_Q S/(\pi v_F J_{int})$, where $G_Q = \frac{\pi}{\hbar} n_{imp}^{2D} (e v_F J_{int})^2$ and $S = 7/2$. Thus, $G_r/G_i = 1/(\pi S v_F J_{int})$, and we are able to reach the case of $G_i > G_r$ in the limit of $v_F J_{int} \ll 1$, which is our case ($v_F J_{int} \approx 0.02$).

A recent work based on spin sinking in lateral spin valves estimates $|G_s| \approx 3 \times 10^{12}$ $\Omega^{-1} m^{-2}$ in EuS/Cu at 10 K (ref 57), with EuS being also evaporated on top, in good agreement with our value of $|G_s| \approx 1.1 \times 10^{12}$ $\Omega^{-1} m^{-2}$ at the same temperature. Another work on EuS evaporated on graphene estimates an interfacial exchange field >14 T based on measurements of the Zeeman spin Hall effect at 4.2 K (ref [42]). We can calculate the interfacial exchange field $h_{ex}$ that is related to $G_i$ obtained from our SMR measurements by using the following expression[54]: $h_{ex} = G_i / \pi G_0 v_F b$, where $G_0$ is the conductance quantum and $b$ is a length of the order of the mean free path ($l$). From the value $G_i \approx 7.0 \times 10^{12}$ $\Omega^{-1} m^{-2}$ observed at the lowest temperature, and taking typical values $v_F \approx 3-4 \times 10^{28}$ $m^{-3} eV^{-1}$ (ref 74) for a metal and $l \approx 10^{-9}$ m for highly resistive Pt, we obtain $h_{ex} = 0.72-0.96$ meV (equivalent to 12.4–16.6 T), in good agreement with ref 52.

As mentioned in the introduction, the interfacial exchange field plays a crucial role in several applications. Therefore, an accurate control of it is very important. For example, for bolometers and cryogenic memories[46,47] based on the spin-splitting induced in EuS/Al bilayers, the effective spin-splitting field induced in Al is given by $h_{eff} = h_{ex} b / d_{Al}$, where $d_{Al}$ is the thickness of the Al layer. If we assume the same order of magnitude of the interfacial $h_{ex}$ that we obtain for Pt, for $d_{Al} \approx 3-10$ nm (refs 42 and 51) we obtain $h_{eff} \approx 0.07-0.3$ meV, whereas the superconducting gap for Al at low temperatures is approximately $\Delta \approx 0.2$ meV (ref 42). In order to observe coexistence between superconductivity and the spin-splitting field, the effective exchange field may not exceed the paramagnetic limit, which at low temperatures is $h_{eff} < 0.7\Delta \approx 0.14$ meV (ref 75). Hence, for the observation of a clear spin-split Bardeen–Cooper–Schrieffer (BCS) density of states, special care should be taken in the fabrication of EuS/Al bilayers.

In summary, we characterize the spin transport in a Pt/EuS interface, where EuS is a pure ferromagnetic insulator below ~19 K, by using spin Hall magnetoresistance. We observe a substantial anomalous Hall-like contribution of the SMR, driven by a large imaginary part of the spin-mixing conductance. We apply a microscopic theory of SMR to extract relevant parameters such as the exchange interaction between the conduction electrons in Pt and the localized magnetic moments in EuS ($J_{int}/a^3 \sim 3-4$ meV). We also obtain the temperature dependence of the interfacial spin conductances, experimentally demonstrating a larger field-like torque ($G_i$) than damping-like torque ($G_r$) in a heavy metal/magnetic insulator interface. The strong interfacial exchange field associated to $G_i$ is estimated as 0.72–0.96 meV (12.4–16.6 T). Therefore, SMR measurements offer a simple way to quantify effective exchange fields which are of interest in different areas of Condensed Matter Physics, such as proximity effects in superconducting hybrid systems.

**Methods.** *Device fabrication.* EuS/Pt samples were prepared by patterning a Pt Hall bar (width $w$=500 μm, length $L$=900 μm, and thickness $d_N$=5 nm) on top of a SiO$_2$(150 nm)/Si substrate by photolithography process and magnetron-sputtering deposition (80 W; 3 mTorr). A EuS layer was *ex-situ* evaporated on top of the Pt film. For growing the EuS, the patterned sample was inserted in a UHV preparation chamber (base pressure 5×10$^{-9}$ mbar) and left for twelve hours at room temperature to remove the water absorbed at the Pt surface. EuS was grown by means of sublimation of a stoichiometric EuS powder (99.9% purity) in a commercial e-beam evaporator. During preparation the substrate was kept at room temperature and the growth rate was calibrated with a quartz microbalance (0.5 nm/min). The total thickness of the EuS layer is 14 nm.

*Materials Characterization.* The magnetic properties of the EuS film were studied by vibrating sample magnetometer (VSM). The quality of the Pt/EuS interface by (scanning) transmission electron microscopy (S)TEM, performed on a TitanG2 60–300 electron microscope (FEI Co., The Netherlands). The composition profiles were acquired in STEM mode utilizing energy dispersive X-ray spectroscopy (EDX) signal.

*Electrical measurements.* I-V curves of the EuS film were performed in a variable-temperature probe-station (Lakeshore) under high vacuum, by using a Keithley 4200 semiconductor analyzer. Magnetotransport measurements in Pt were carried out in a liquid-He cryostat at temperatures $T$ between 2.5 K and 40 K, externally applied magnetic fields $\mu_0 H$ up to 4 T, and a 360º sample rotation, by using a "DC reversal" technique with a Keithley 2182 nanovoltmeter and 6221 current source.

▪ **ASSOCIATED CONTENT**

**Supporting information**

Additional details on the electrical, magnetic and structural properties of the EuS films, disentangling magnetoresistance effects from the ADMR and FDMR measurements, and the magnetic field sweep history observed in low-field behavior of FDMR measurements.

▪ **AUTHOR INFORMATION**

**Corresponding Author**
*E-mail: f.casanova@nanogune.eu

**Notes**
The authors declare no competing financial interest.

▪ **ACKNOWLEDGMENTS**

The authors thank the technical support provided by SGIker Medidas Magneticas Gipuzkoa (UPV/EHU/ERDF, EU). The work was supported by the Spanish MICINN under the Maria de Maeztu Units of Excellence Programme (MDM-2016-0618), and Project Nos. MAT2015-65159-R, MAT2016-78293-C6-5-R, FIS2017-82804-P, and RTI2018-094861-B-100, by the Regional

Council of Gipuzkoa (Projects Nos. 100/16 and IT-1255-19), by the EU's Horizon 2020 research and innovation program under Grant Agreement Nos. 800923-SUPERTED and 766025-QuESTech. J.M.G.-P. thanks the Spanish MICINN for a Ph.D. fellowship (Grant No. BES-2016-077301). M.G. acknowledges support from the European Commission for a Marie Sklodowska-Curie IEF (Grant No. 748971-SUPER2D) and from la Caixa Foundation for a Junior Leader fellowship (Grant No. LCF/BQ/PI19/11690017).

- **REFERENCES**

# Supporting information

## S1. Electrical characterization of the EuS thin film

We characterized the electrical properties of a 14-nm-thick film of EuS on top of a $SiO_2/Si$ substrate which was deposited simultaneously with the sample containing the Pt Hall bar used for the SMR measurements in the main text. Due to the large resistance of the film, we performed 2-point measurements by applying a fixed voltage and measuring the current. With this, we can obtain *I-V* curves from where the resistance can be estimated (see Figure S1). We measured a resistance of ~68 G$\Omega$ at 300 K (Figure S1a), which increases up to ~10 T$\Omega$ at 200 K (Figure S1b). Below 200 K, we are not able to measure the resistance of the EuS film as it becomes larger than the internal impedance of our instrument (see Figure S1c at 120 K). Therefore, we can confirm that EuS is an insulator and all the charge current flows into the Pt Hall bar.

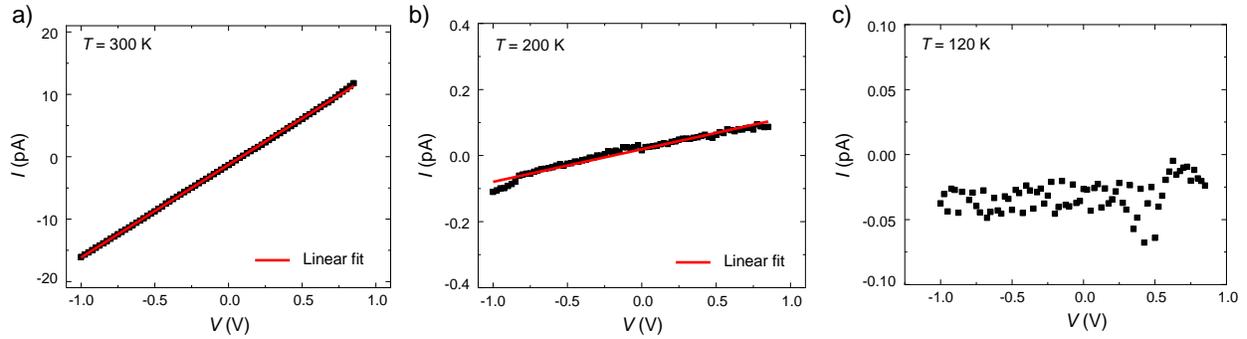

Figure S1. Current-Voltage (*I-V*) curves at (a) 300 K, (b) 200 K and (c) 120 K in a 14-nm-thick film of EuS evaporated on $SiO_2/Si$. The red lines are the linear fit to extract the EuS resistance.

## S2. Influence of $Eu^{2+}$ and $Eu^{3+}$ to the saturation magnetization of the EuS thin film

The saturation magnetization at the lowest temperature (~ 500 kA/m, corresponding to ~3 $\mu_B$/Eu atom) is below the bulk value (1240 kA/m, 6.9 $\mu_B$/Eu atom).[1] This difference is expected in thin films as compared to bulk material.[2] This discrepancy is usually ascribed to two main factors. First, a fraction of the Eu in evaporated EuS thin films unavoidably possess the $Eu^{3+}$ hybridization instead of the expected $Eu^{2+}$. In pure, stoichiometric EuS, the Eu ion is in a 2+ charge state; the $Eu^{3+}$ indicates the presence of paramagnetic $Eu_3S_4^+$, which decreases the saturation magnetization. Second, exchange coupling in EuS is direct and depends strongly on the defects, grain boundaries, etc.[3] As a result, even for a stoichiometric EuS, defects and grain boundaries decrease the saturation magnetization.

In our samples, we measured a saturation magnetization which is approximately half as compared to the reported value for bulk films. To account for such difference, which was previously observed in thin films, we estimated the relative amount of the $Eu^{3+}$ and $Eu^{2+}$ ions in our EuS film through x-ray photoelectron spectrometry (XPS). Figure S2 shows the XPS spectrum of a EuS film grown on Si with the same conditions as the EuS film used in the main text. From this spectrum, we can estimate that the amount of $Eu^{2+}$ is about 80% of the total Eu. Such relative amount of $Eu^{2+}$ is relatively high for an evaporated film, indicating the high quality of our film.[4] Based on this proportion, one could expect a saturation magnetization close to 80% of its bulk value, which is not our case. However, our film is polycrystalline, characterized by grains with a diameter of few nm (as shown by the TEM characterization). Therefore, the number of grain boundaries/defects in the phase containing $Eu^{2+}$ accounts for the remaining difference in the saturation magnetization. This behavior is in a very good agreement with ref 2.

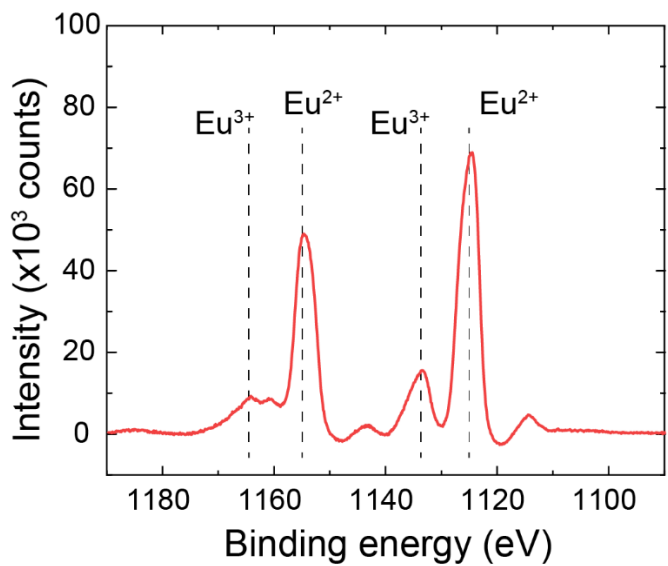

Figure S2. XPS spectrum for a EuS film grown with the same conditions as the one presented in the main text.

## S3. Composition and crystallographic structure of the EuS thin film

On the one hand, the curves on Figure 1b of the main text do not represent the atomic percentage neither the raw data from the Energy dispersive X-ray spectroscopy (EDX) measurements. They are the result of multiple linear least squares (MLLS) analysis of spectral data accounting for the cross-sections for X-ray excitation of all lines for all elements. This is not a quantitative data in the sense that EDX is not that precise to give an exact composition at least within ±10% error. However, even if it is a qualitative representation, we can say the EuS film has the approximately right composition.

On the other hand, the crystallographic structure of the EuS film evaporated on top of a polycrystalline Pt thin film can be obtained from the HR-TEM image in Figure 1b of the main text. Figure S3a shows the fast Fourier transform (FFT) of the HR-TEM image. The FFT pattern shows bright spots that represent the different crystal orientations of the EuS film, which is polycrystalline. We can perform a one-dimensional average of the FFT pattern to obtain the crystal direction spectrum, which is plotted in Figure S3b. Figure S3b shows the different intensity peaks for the crystal orientations present in our EuS film and, in dashed lines, we place the different peaks for the face centered cubic (fcc) structure of bulk EuS. We can confirm the fcc structure is maintained in our thin film, with a strong texture along the (200) orientation in the direction normal to the surface. From the HR-TEM image in Figure 1b of the main text, we can estimate a grain size of few nm.

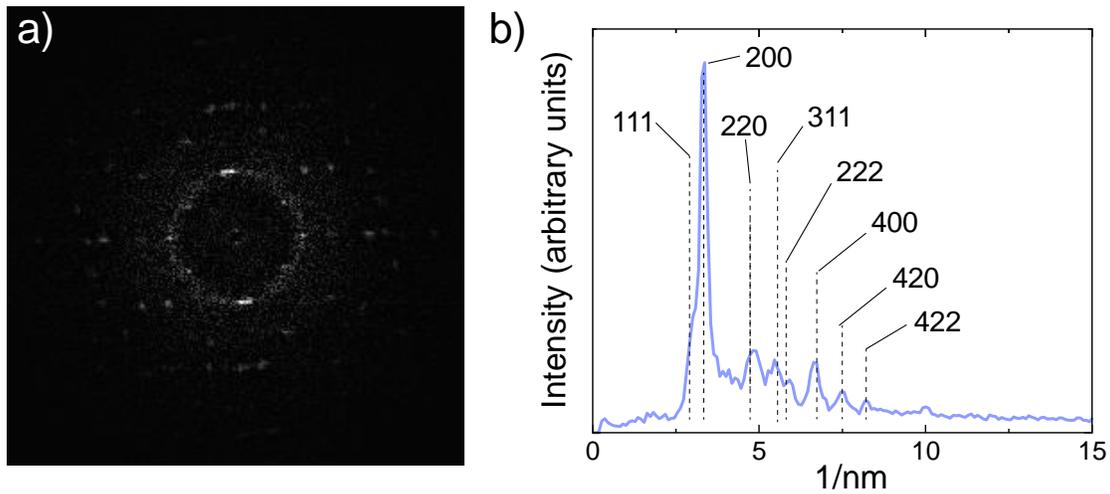

Figure S3. (a) Fast Fourier transform (FFT) of the HR-TEM shown in Figure 1b of the main text. (b) One-dimensional average of the FFT pattern (blue solid line). The black dashed lines are the representative peaks for a EuS fcc crystallographic structure.

**S4. Disentangling magnetoresistance effects from the ADMR and FDMR measurements**

When Pt is in contact with a ferromagnet (FM), it can become ferromagnetic via magnetic proximity effect (MPE), showing anisotropic magnetoresistance (AMR) and anomalous Hall effect (AHE) that would mirror the magnetization of the adjacent FM. However, SMR also mirrors the magnetization of the adjacent FM.

It is possible to distinguish the SMR scenario from the MPE scenario by performing the typical angular-dependent magnetoresistance (ADMR) measurements in the three main H-rotating planes ($\alpha$, $\beta$ and $\gamma$), because the SMR symmetries are clearly different than those of the AMR associated to MPE.[6] Whereas a $cos^2$ modulation in $\alpha$ and $\beta$ is expected for SMR, a $cos^2$ modulation in $\alpha$ and $\gamma$ is expected for AMR. Although the ADMR results in Figure 1d of the main text mostly show the SMR symmetry, there is a small modulation in the $\gamma$–plane at 2 T in our EuS/Pt system. In addition to AMR, other magnetoresistances such as weak antilocalization (WAL)[6,7] and ordinary magnetoresistance (OMR)[8] can lead to a modulation in $\gamma$ at moderate fields. If this modulation is present, the origin can be identified by looking at the field-dependent magnetoresistance (FDMR) measurements, which will be different for AMR, WAL and OMR.

A MPE origin would show, after magnetization reversal, the same saturation baseline for FDMR($H_z$) and FDMR($H_y$) and a different one for FDMR($H_x$). However, what is observed in Figure 2a is the same baseline for FDMR($H_z$) and FDMR($H_x$) and a different one for FDMR($H_y$), which is the expected behavior for SMR. This result rules out AMR and confirms SMR as the magnetoresistance at low fields.

The origin of the $\gamma$–plane modulation at 2 T can thus be identified from the high-field behavior of the FDMR curves plotted in Figure 2a. OMR is relevant only when the resistivity is low enough,[8] which is not the case of the present Pt. WAL, instead, is only present in disordered Pt at low temperatures, showing an out-of-plane MR with a characteristic inverted bell-shaped curve in $H_z$ direction, and also a smaller, parabolic-like MR when the field is in plane ($H_x$ and $H_y$ directions).[6,7] Although this explains the $\gamma$–plane modulation at 2 T, WAL behavior alone cannot explain the curves in Figure 2a where, at high fields, FDMR($H_z$) > FDMR($H_x$) > FDMR($H_y$).

In order to understand this result, we also have to take into account Hanle magnetoresistance (HMR), which has the same origin as SMR and appears when the magnetic field is applied perpendicular to the spin accumulation generated in Pt due to the induced precession, leading to the same parabolic-like behavior in $H_x$ and $H_z$ direction and no variation in $H_y$ direction.[7] HMR only depends on the nature of the heavy metal and shows a weak variation with temperature between low and room temperature. Note that HMR, as SMR, will not contribute to any $\gamma$–plane modulation.

The behavior of the FDMR curves in Figure 2a can thus be fully explained with the combination of HMR and WAL (in addition to the SMR, which is most evident at low magnetic fields). This leads to the observed shape of the 3 curves and the order FDMR($H_z$) > FDMR($H_x$) > FDMR($H_y$). The very same behavior is observed in YIG/Pt and reported in ref 7 [Figure S4(a) of its Supplemental Material].

**S5. Magnetic field sweep history observed in low-field behavior of FDMR measurements**

The states of the FDMR curves at $H_x$=0, $H_y$=0, and $H_z$=0 shown in Figure 2a,b of the main text can be simply explained with the magnetic field sweep history. Whereas the difference in the FDMR curve for $H_x$ and $H_y$ at zero field is explained by the remanence, which keeps the magnetization in the x- and y-direction, respectively, in the case of $H_z$ the shape anisotropy prevents the magnetization to point out of plane at zero field. From Figure 2a,b of the main text, the magnetization seems to fall towards the x-direction when $H_z$ crosses zero, as the resistance state for $H_x$=0 and $H_z$=0 is the same within the noise level.

In order to understand the low-field behavior of the FDMR curve for $H_z$, we show the results of the FDMR curves measured in a sister sample. Although the sample is different from the one in the main text, it shows the same global behavior. Interestingly, in this sample, the FDMR curves where measured in three different sample holder setups, to allow the sample to rotate in the $\alpha$–plane, $\beta$–plane or $\gamma$–plane. At each configuration, the FDMR curve can be measured with the magnetic field in two of the three main axes. Figure S4 shows all six FDMR curves, two for $H_x$, two for $H_y$, and two for $H_z$. Whereas the FDMR curves for $H_x$ and $H_y$ are the same regardless of the sample holder setup (as expected), the two FDMR curves for $H_z$ behave differently below the out-of-plane saturation field of ~1.5 T. The FDMR($H_z$) curve in the $\beta$–plane configuration has the expected shape, following the non-hysteretic, continuous magnetization rotation, with a resistance value at zero field half way between the FDMR($H_x$) and FDMR($H_y$) curves, corresponding to a zero net magnetization (also in agreement with Figure 2d). In contrast, the FDMR($H_z$) curve in the $\gamma$–plane configuration has a more complex behavior, with hysteretic magnetization reversal peaks. Although the two measurements are nominally the same, the difference observed below ~1.5 T strongly supports a sample misalignment in the $\gamma$–plane. This scenario will favor a net magnetization in the x-direction at low fields due to the presence of an x-component of the magnetic field when $H_z$ is swept through zero, explaining the presence of hysteretic magnetization reversal peaks in this sample, and the resistance state for $H_x=0$ and $H_z=0$ being the same within the noise level in the sample of the main text.

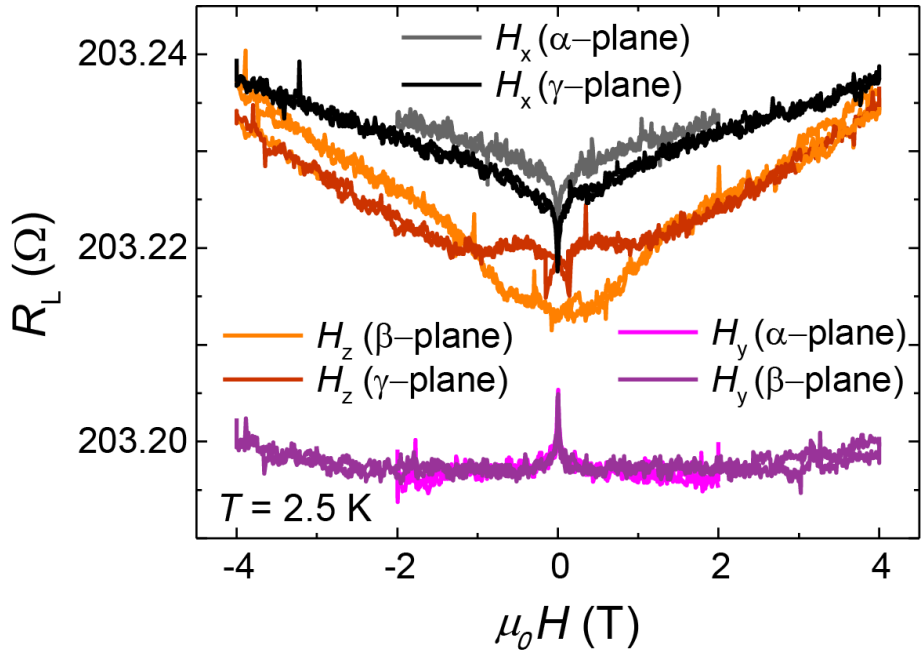

Figure S4. Longitudinal FDMR measurements performed along the three main axes at 2.5 K. Three different sample holder setups are used ($\alpha$–, $\beta$– or $\gamma$–plane), each one accessing two of the three magnetic field directions. This way, the FDMR curve for each magnetic field direction is measured twice.